\documentstyle[12pt]{article}
\begin{document}
\date{}

\title{Instability of the planetary orbits for space-time dimensions higher than four}
\author{R. D. Mota  and A. P\'erez-Guerrero} \maketitle
\begin{minipage}{0.9\textwidth}
\begin{center}
{\it Unidad Profesional Interdisciplinaria de Ingenier\'{\i}a y
Tecnolog\'{\i}as Avanzadas, Instituto Polit\'ecnico Nacional, M\'exico}
\end{center}
Av. IPN No. 2580, Col. La Laguna Ticom\'an, Delegaci\'on Gustavo
A. Madero, 07340 M\'exico D. F., M\'exico.
\end{minipage}
\begin{abstract}
It is shown that in a Minkowski space of total space-time
dimension $D=d+1$, the orbits of the planetary motion are stable
only if the total dimension of space-time is $D\le 4$. The
proof is performed in a fully didactic way.
\end{abstract}
\section{INTRODUCTION}
Bertrand's theorem  establishes that the only central potentials
with stability are the harmonic oscillator and the Coulomb's one
\cite{GOLD}. It has been related the closeness of the classical
orbits to their symmetries \cite{MEXI}. The existence of different
kinds of factorization operators for the radial Schr\"odinger
equation corresponding to the Kepler and harmonic oscillator problems has been
attributed to the classical stability of these systems according
to the Bertrand's theorem \cite{CHINOS1}. Also, a generalization of this
result has been developed when the potential has suitable extra
angular momentum \cite{CHINOS2}. More subtle points has been
studied like the connection between closed orbits and their stable
symmetries for small perturbations \cite{RUSOS}. It has been found
that when all eight symmetries are stable the orbits are closed,
but not necessarily if only three symmetries of them are stable,
as it happens in the relativistic case where the perihelion of the
orbit advances \cite{RUSOS}.

Recently it has been studied the circular orbits stability for the
Schwarzs- child geometry ($D=4$) in commutative and noncommutative
spaces \cite{IRAN}. Schwarzschild solution in commutative space possesses
stable circular orbits for a radial distance $r> 6GM$ and unstable
ones for $3GM < r < 6GM$, where $M$ is the mass of the attractor
center and $G$ is the gravitational constant. The space
non-commutativity increases the radius of stable circular orbits
\cite{IRAN}. Gravitational classical instability in higher
dimensions ($D>4$) has been explored for various static and non
static spacetimes: black holes, brane worlds, generalized
Schwarzschild, AdS black strings, "bubbles of nothing"
\cite{GIBBONS} and hyperbolic spaces \cite{NEUPANE}.

The aim of the present paper is to show that the planetary circular
orbits in a Minkowski spacetime of total dimension $D=d+1$, are
stable only when $D\le 4$. To obtain this result the Bertrand's
theorem plays a central role which allow us to give an elegant and
didactic proof. In section 2, we obtain the gravitational force in
$d$ spacial dimensions due to a particle of mass $M$. In section
3, we sketch the Bertrand's theorem, and it is applied to constraint 
the spatial dimensions in order to obtain the
stability of circular orbits. In section 4, we give the concluding
remarks.

\section{GRAVITATIONAL FORCE IN A \\$d$-DIMENSIONAL SPACE}

The magnitude of the attractive gravitational force between two
point particles $m$  and $M$ in the three-dimensional space is
given by
\begin{equation}
\vec F=-G^{(3)}\frac{mM}{r^2}\hat r,\label{3dgra}
\end{equation}
which satisfies the Gauss' law (setting $m=1$) in three dimensions
\begin{equation}
\int_{S^2(r)} {\vec F}\cdot d{\vec a}=-4\pi
G^{(3)}M_{enc}\label{gauss},
\end{equation}
where $M_{enc}$ is the enclosed  mass by the Gaussian sphere $S^2$
and $G^{(3)}\equiv G$.

We are interested in generalizing expression (\ref{3dgra}) to
a $d$-dimensional space. In order to obtain this, we demand that the
gravitational force satisfies the Gauss' law (\ref{gauss}) for any
$d$ value, remaining fixed the value of the right hand side. This
means that the flux of the gravitational force on any sphere
$S^{d-1}(r)$ is always equal to the enclosed mass times $4\pi
G^{(3)}$ \cite{STEFAN}.

Some remarks about $d\vec a$ are necessary. If we consider the
spheres $S^1$ and $S^2$, their areas are equal to $2\pi r$ and
$4\pi r^2$, respectively. The area of the sphere $S^3$ is some
alike a three-dimensional volume. To higher-dimensional spaces
the area of a sphere $S^{d-1}$ of radius $r$ is known  as the
volume,  $a\equiv$(vol$(S^{d-1}(r))$). In what follows we use this
terminology to denote areas for any $d$ value. Hence, the volumes
for $S^1$ and $S^2$ are
\begin{eqnarray}
&&\hbox{vol}(S^1(r))=2\pi r,\label{vol1}\\
&&\hbox{vol}(S^2(r))=4\pi r^2\label{vol2}.
\end{eqnarray}
Thus, the factor $4 \pi$ in equation (\ref{gauss}) is equal to the
volume of the unit sphere $S^{2}$. Equations (\ref{vol1}) and (\ref{vol2}), 
lead to infer that
for a sphere in a $d$-dimensional space the equality
vol$(S^{(d-1)}(r))=r^{d-1}$vol$(S^{d-1})$ must hold, being
vol$(S^{d-1})$ the volume of the unit sphere. Moreover, 
the volume differentials ($r$
fixed) are
\begin{eqnarray}
&&d[\hbox{vol}(S^1(r))]=r d\theta=rd[\hbox{vol}(S^1)],\\
&&d[\hbox{vol}(S^2(r))]=r^2\sin(\theta)d\theta
d\phi=r^2d[\hbox{vol}(S^2)],
\end{eqnarray}
from which we conclude that in general, for a $d$-dimensional
sphere, the equality
\begin{equation}
da=d[\hbox{vol}(S^{d-1}(r))]=r^{d-1}d[\hbox{vol}(S^{d-1})]\label{da}
\end{equation}
holds.

In order
to the gravitational force flux on the sphere $S^{d-1}(r)$ be the
fixed constant $4\pi G^{(3)}$, according to the result (\ref{da}), 
the force between two massive point particles 
must depend on $1/r^{d-1}$. We propose
it has the form
\begin{equation}
\vec F=-G^{(d)}\frac{mM}{r^{d-1}}\hat r.
\end{equation}
Now we focus our attention on the right hand side of equation
(\ref{gauss}). The mass $M$ as an intrinsic matter property must
not depend on the space dimension. Hence, the product $4\pi
G^{(3)}=\hbox{vol}(S^2)G^{(3)}$ must have a fixed value for any
$d$. Since in higher dimensions the area (volume) of the unit sphere
depends on the space dimension, the gravitational constant must
{\it depend} on the dimension too. Hence,
$\hbox{vol}(S^{d-1})G^{(d)}=4\pi G^{(3)}$. Above remarks allows to
write the generalized Gauss' law as
\begin{equation}
\int_{S^{d-1}(r)} {\vec F}\cdot d{\vec a}=-\hbox{vol}(S^{d-1})
G^{(d)} M_{enc}\label{gauss2}.
\end{equation}
which is satisfied by the gravitational force
\begin{equation}
\vec F=-\frac{4\pi
G^{(3)}}{\hbox{vol}(S^{d-1})}\frac{mM}{r^{d-1}}\hat
r,\label{ddgra}
\end{equation}
for $m=1$.

The next step is to find the volume of a unit sphere of dimension
$d-1$. We proceed  as in \cite{ZWIE} and calculate the integral
\begin{equation}
I_d=\int_{-\infty}^{\infty}\int_{-\infty}^{\infty}
\cdot \cdot \cdot \int_{-\infty}^{\infty}dx_1 dx_2\cdot \cdot \cdot dx_d
e^{-r^2}=\int_{R^d}e^{-r^2}dv\label{INT}
\end{equation}
in {\it all space} in two ways.

First, since $r^2=x_1^2+x_2^2+...+x_d^2$, the integrand of
(\ref{INT}) is separable and it reduces to a product of
one-dimensional integrals. Thus,
\begin{equation}
I_d=\prod_{i=1}^d \int_{-\infty}^{\infty} dx_i e^{-x_i^2}\label{int}
\end{equation}
However, each integral of the product is well known from the
probability theory, it is $\int_{-\infty}^\infty dx_i
e^{-x_i^2}=\pi^{\frac{1}{2}}$. Therefore, $I_d=\pi^\frac{d}{2}$.

Second, the value of $I_d$ is obtained by considering the space
$R^d$ as divided by thin spherical shells. At fixed $r$ the space
is the sphere $S^{d-1}(r)$ and the volume of the space between $r$
and $r+dr$ is $dv=$vol$(S^{d-1}(r))$$dr$. This fact allows to write 
equation (\ref{INT}) as 
\begin{eqnarray}
I_d&=&\int_0^\infty
\hbox{vol}(S^{d-1}(r))e^{-r^2}dr\\
&=&\hbox{vol}(S^{d-1})\int_0^\infty r^{d-1}e^{-r^2}dr\\
&=&\frac{1}{2}\hbox{vol}(S^{d-1})\int_0^\infty
t^{\frac{d}{2}-1}e^{-t}dt,
\end{eqnarray}
where we have used the relation
vol$(S^{(d-1)}(r))=r^{d-1}$vol$(S^{d-1})$ and  performed the
variable change $t=r^2$. The last integral is identified with the
integral representation of the gamma function defined by
$\Gamma(x)\equiv \int_0^\infty t^{x-1}e^{-t}dt$, with $x>0$. With
this result we obtain that
$I_d=\frac{1}{2}\hbox{vol}(S^{d-1})\Gamma(\frac{d}{2})$. Thus by
equating our two results for $I_d$, we obtain
\begin{equation}
\hbox{vol}(S^{d-1})=\frac{2\pi^{\frac{d}{2}}}{\Gamma(\frac{d}{2})}
\label{dunitsphere}.
\end{equation}
Results (\ref{ddgra}) and (\ref{dunitsphere}) complete our purpose
to find the gravitational force between two point particles in a
$d$-dimensional space.

Bertrand's theorem ensures that the circular planetary orbits
around the sun in our four-dimensional space-time are stable under
small perturbations \cite{GOLD}, and relates its stability to the
values of the power in potentials with the form $V=r^\nu$. We
notice that the force (\ref{ddgra}) due to one particle of mass
$M$ on a particle of mass $m$ is proportional to $1/r^{d-1}$. This
fact will allow us to apply Bertrand's theorem to find the values
of the spatial dimension $d$ in order to have stable planetary
orbits in higher dimensions.

\section{THE BERTRAND'S THEOREM AND \\STABILITY OF THE CIRCULAR 
ORBITS \\IN A $d$-DIMENSIONAL  SPACE}
We begin by reviewing the Bertrand's theorem proof.  
Since the total angular momentum is conserved, this allows us to
set an inertial frame in the mass center of the system. Also,
since the force is central, the angular momentum is conserved and
restrain the orbit to be plane. Hence, the motion of the two
particles is reduced to that of a single one with reduced mass
$\mu=\frac{mM}{m+M}$ under the potential $V(r)$, being $r$ the
distance between particles. It is usual to set the $z$-axis in the
angular momentum direction, orthogonal to the orbit plane. Also,
since  $V(r)$ does not depend on time $t$, the total energy
$E$ is conserved.

Since the lagrangian of the system is
\begin{equation}
L=T-V=\frac{1}{2}\mu({\dot r}^2+r^2{\dot\theta}^2)-V(r)
\end{equation}
and does not depend on the variable $\theta$, the angular momentum
$\ell=\mu r^2\dot\theta $ is conserved. The total energy of the
system is
\begin{eqnarray}
E&=&\frac{1}{2}\mu({\dot r}^2+r^2{\dot\theta}^2)+V(r)\\
&=&\frac{1}{2}\mu \dot r^2+\frac{1}{2}\frac{\ell^2}{\mu r^2}+V(r)
\\ &\equiv &\frac{1}{2}\mu \dot r^2+ V_{ef},
\end{eqnarray}
from which is obvious the definition of the effective potential
$V_{ef}(r)$. For the case in our study, the reduced mass is $
\mu=\frac{mM}{m+M}\approx m$.  Thus, it follows that
\begin{equation}
\frac{\partial V_{ef}}{\partial r}=\frac{\partial V(r)}{
\partial r}-\frac{\ell ^2}{mr^3}\label{pot},
\end{equation}
or
\begin{equation}
F_{ef}=F+\frac{\ell ^2}{mr^3},\label{Fefec}
\end{equation}
where $F\equiv -\frac{\partial V}{\partial r}$.

The condition to have circular orbits is imposed by $\dot r =0$,
and they are stable only when the potential has an effective
minimum at a distance $r=r_0$. This fact implies that for circular
orbits, $F_{ef}=-\left. \frac{\partial V_{ef}}{\partial r}
\right|_{r=r_0}$ must be equal to zero (whenever the case, a
maximum or a minimum). Thus, equation (\ref{Fefec}) reduces to
\begin{equation}
F=-\frac{\ell ^2}{mr_0^3}.
\end{equation}

By demanding that the effective potential $V_{ef}(r)$ get a
minimum at $r=r_0$, from equation (\ref{pot}) it must satisfy
\begin{equation}
\left. \frac{\partial^2 V_{ef}}{\partial r^2}
\right|_{r=r_0}=\left. \frac{\partial^2V}{\partial r^2}
\right|_{r=r_0}+3\frac{\ell ^2}{m{r_0}^4}>0
\end{equation}
this implies
\begin{equation}
\left. -\frac{\partial F}{\partial r}\right|_{r=r_0}>-3\frac{\ell
^2}{m{r_0}^4}=\frac{3}{r_0}F(r)|_{r=r_0},
\end{equation}
or
\begin{equation}
\left. \frac{\partial F}{\partial
r}\right|_{r=r_0}<-\frac{3}{r_0}F(r_0).\label{ineq}
\end{equation}
This inequality must be satisfied in order to have stable circular orbits at $r=r_0$.

From equation (\ref{ddgra}) and (\ref{dunitsphere}), the force between two particles in
a $d$-dimensional space is given by
\begin{equation}
F=-\frac{4\pi
G^{(3)}}{\hbox{vol}(S^{d-1})}\frac{mM}{r^{d-1}}=-
\frac{2 \Gamma(\frac{d}{2})G^{(3)}}{\pi^{\frac{d}{2}-1}}\frac{mM}{r^{d-1}}\label{efec}.
\end{equation}
Hence, 
\begin{eqnarray}
\frac{d F}{dr}
=\frac{2\Gamma(\frac{d}{2})}{\pi^{\frac{d}{2}-1}}G^{(3)}M
 m(d-1)r^{-d}\label{difefe}.
\end{eqnarray}
By substituting  equations (\ref{efec}) and (\ref{difefe})
evaluated at $r=r_0$ into equation (\ref{ineq}), we obtain
\begin{equation}
\frac{d-1}{r_0^d}<\frac{3}{r_0^d},
\end{equation}
or $d<4$. This means that the maximum value of the spatial
dimension $d$ is 3, and it is the upper bound we were looking for.

\section{CONCLUDING REMARKS}
We have shown that the spatial dimension must satisfy $d<4$ to
have stable circular orbits. This means that for the planetary motion
in a space-time of five dimensions or higher, the circular orbits
become instable. This result could be derived because the
gravitational force in $d$ dimensions possesses the same
mathematical form as those for the central potentials studied by
Bertrand. Although the problem we have treated in this paper is a
texbook one \cite{ZWIE}, as we have shown above, its solution is
non-trivial. Moreover, as was emphasized in the
introduction, the concept of stability is studied in many current
research areas of physics. Being this paper a comprehensive and
fully detailed work, it can be a helpful pledge for the students
to make a glance for advanced topics.

\end{document}